\documentclass{ws-ijmpa}
\usepackage[super,compress]{cite}
\usepackage{graphicx}
\begin{document}

\markboth{V.~M.~Mostepanenko}
{Some remarks on axion dark matter and dark energy}

\catchline{}{}{}{}{}

\title{Some remarks on axion dark matter, dark energy and energy of the quantum vacuum}

\author{ V.~M.~Mostepanenko}

\address{Central Astronomical Observatory at Pulkovo of the
Russian Academy of Sciences, Saint Petersburg,
196140, Russia\\
Institute of Physics, Nanotechnology and
Telecommunications, Peter the Great Saint Petersburg
Polytechnic University, Saint Petersburg, 195251, Russia\\
Kazan Federal University, Kazan, 420008, Russia\\
vmostepa@gmail.com}

\maketitle

\begin{history}
Received 8 August 2019
\end{history}

\begin{abstract}
In connection with the problem of dark matter, we discuss recent results on
constraining the parameters of axion-to-nucleon interaction following from
the experiment on measuring the difference of Casimir forces. It is shown
that this experiment not only leads to competitive constraints, but provides
stronger support to other constraints obtained in Casimir physics so far.
The description of dark energy by means of cosmological constant originated
from the quantum vacuum is considered in terms of the renormalization
procedures in quantum field theory. It is argued that only the renormalized
value of cosmological constant directly connected with the observed density
of dark energy is of physical significance, so that some statements in the
literature concerning the vacuum catastrophe may be considered as an
exaggeration.
\keywords{Axions; dark matter; dark energy; cosmological constant.}
\end{abstract}

\ccode{PACS numbers.: 14.80.Mz, 95.35.+d, 95.36.+x, 98.80.Es }

\section{Introduction}

At the moment the situation in fundamental physics is much in common with that
in the end of the nineteenth century. At that time it was commonly believed
that all space is full of hypothetical substance, ether, needed to ensure
propagation of the electromagnetic waves. According to current concept, more
than 95\% of the energy of the Universe consists of the so-called dark energy
and dark matter which we are presently not capable to observe directly.\cite{1}
The single difference is that we already have several indirect evidences for
the existence of dark energy and dark matter and hope to get the direct ones in
the immediate future, whereas all attempts to find some experimental fact in
favor of ether more than a centure ago failed.

It has long been known that stellar motion in the neighborhood of our galaxy
cannot be explained by the gravitational theory if the mass of galaxy is not
much larger than that of visible matter. The same is true for the clusters of
galaxies.\cite{2} The missing dark matter contributes up to 27\% of the
energy of our Universe. There are many attempts to understand the nature of
dark matter, but the most realistic model suggests that it consists of light
uncharged pseudoscalar particles, axions, whose interaction with familiar
elementary particles, such as photons, electrons and nucleons, is very
weak.\cite{3}.

One more unseen substance is the dark energy which was introduced rather
recently in order to explain the accelerated expansion of the Universe in
the framework of General Relativity Theory.\cite{4}. Unlike the dark matter,
the dark energy is characterized by the negative pressure which makes
possible the effect of accelerated expansion. The dark energy contributes
approximately 68\% of the Universe energy. The most plausible explanation
for the dark energy is by means of the cosmological constant in Einstein's
equations which can be interpreted as originated from the quantum vacuum.
In doing so, however, the so-called vacuum catastrophe arises\cite{5}
because the value of the cosmological constant predicted by quantum field
theory greatly exceeds the one needed to explain an observed acceleration
of the Universe expansion.

In this paper, we discuss new constraints on the axion-to-nucleon coupling
constant obtained from measuring the difference of Casimir forces, as well
as the possibilities to obtain even stronger constraints by optimizing
several other experiments of Casimir physics. We also consider the problem
of dark energy and the possibility of its resolution by means of the
cosmological constant originated from the quantum vacuum. It is argued
that with a proper renormalization procedure only the energy density
corresponding to a physical value of the cosmological constant should be
considered as the source of gravitational interaction which settles the
problem of vacuum catastrophe.

The paper is organized as follows. In Sec. 2 we discuss axions, axion-like
particles and new constraints on their parameters obtained in the Casimir
physics. Section 3 is devoted to the problem of cosmological constant and
its renormalization. In Sec. 4 the reader will find our conclusions and
a discussion. We use the system of units where $\hbar = c = 1$.

\section{Novel constraints on axion-like particles from Casimir physics}

The search for axion-like particles and, thus, for the axion dark matter is performed
by exploiting their presumed interactions with photons, electrons and nucleons. There is a great number of such kind experiments already performed, in operation and planned using different
laboratory techniques\cite{3} (see also Ref.~\citen{6} for experiments in atomic physics).
Although rather strong constraints are already obtained on the coupling constants of axions
to photons and electrons, the laboratory constraints on axion-to-nucleon interaction remain
relatively weak, whereas the astrophysical constraints suffer from serious uncertainties
in theory of dense nuclear matter.\cite{7,8}

It is well known that interaction of axions with nucleons is described either by pseudoscalar
or pseudovector Lagrangian densities.\cite{9} Both Lagrangians lead to common spin-dependent
effective potential between two nucleons due to exchange of one axion.\cite{10}
After an everaging over the volumes of two unpolarized bodies, this leads to zero force,
i.e., the process of one-axion exchange between nucleons does not manifest itself in
measurements of small forces between two macroscopic bodies. If, however, one considers the
exchange of two axions described by the pseudoscalar Lagrangian, the two nucleons interact
by means of spin-independent effective potential\cite{10}
\begin{equation}
V(r)=-\frac{g_{an}^4}{32\pi^3m_n^2}\,\frac{m_a}{r}\,K_1(2m_ar),
\label{eq1}
\end{equation}
\noindent
where $g_{an}$ is the axion-to-nucleon interaction constant, $m_n$ and $m_a$ are the nucleon
and axion masses, $r$ is the separation distance between two nucleons, and $K_1(z)$ is
the Bessel function of the second kind. It should be particularly emphasized that up to the
present the form of effective potential between two nucleons
due to exchange of two axions, described by the
pseudovector Lagrangian,  remains a mystery.\cite{11}

The effective potential (\ref{eq1}) corresponds to an attractive force. The presence of such
forces can be tested in precise experiments on force measurements between closely spaced
macroscopic bodies and, more specifically, on measuring the Casimir force.\cite{12}
In so doing, the force arising due to two-axion exchange, $F_a(d)$, is calculated in the
experimental configuration by a pairwise summation of potentials (\ref{eq1}) over the volumes
of test bodies with subsequent negative differentiation with respect to separation $d$
between them. Then, the constraints on $g_{an}$ and $m_a$ are obtained from the inequalities
\begin{equation}
|F_a(d)|\leq\Delta F_C(d), \quad
|F_a^{\prime}(d)|\leq\Delta F_C^{\prime}(d),
\label{eq2}
\end{equation}
\noindent
where $\Delta F_C$ or $\Delta F_C^{\prime}$ are the total experimental errors in the
Casimir force or in its gradient according to which quantity is measured.

Following this methodology, the constraints on $g_{an}$ at different $m_a$ were
obtained from experiments on measuring the Casimir-Polder force,\cite{13}
the gradient of the Casimir force,\cite{14} the Casimir pressure,\cite{15}
the lateral Casimir force,\cite{16} and in Ref.~\citen{17} from the Casimir-less
experiment\cite{18} (see Ref.~\citen{19} for a review). From measuring the Casimir
interaction, we eventually reached up to several orders of magnitude stronger
constraints on an axion-to-nucleon interaction constant depending on the
interaction range.

It is the subject of a considerable literature that there is a puzzle in comparison
between the measured Casimir force and theoretical predictions of the fundamental
Lifshitz theory.\cite{12} It was found that for metallic test bodies the theoretical
predictions are consistent with the measurement data only under a condition that
the relaxation properties of free electrons are disregarded in computations
(for a review see Refs.~\citen{12}, \citen{20} and more recent results\cite{21,22,23,24,25,26}).
Taking into account that two theoretical predictions diverged for only a few percent,
some doubts could be casted upon the validity of constraints on an axion\cite{13,14,15,16}
obtained so far from the measure of agreement between the data and one of two predictions.

In this connection, the recent difference force experiment\cite{25} is of great
importance because in its configuration the theoretical predictions made with disregarded
and included relaxation properties of free electrons differ by up to a factor of 1000.
In Ref.~\citen{25} the measured quantity is the difference of Casimir forces between
a Cr and Ni-coated sapphire sphere and Au and Ni sectors of the rotating disc covered
by the homogeneous Ti and Au overlayers. These overlayers greatly enhance the variation
in the difference of Casimir forces when the relaxation properties of free electrons
are disregarded or included in computations. As a result, one theoretical approach,
taking into account the relaxation properties of free electrons, was unambiguously
excluded and another one, disregarding these properties, was conclusively confirmed.

Taking into consideration a distinctive role of the experiment of Ref.~\citen{25},
its measurement results have been used for obtaining constraints on the axion-to-nucleon
coupling constants.\cite{27} The forces $F_a^{\rm Ni,Ni}(d)$ and $F_a^{\rm Ni,Au}(d)$
arising due to two-axion exchange between a Ni-coated sphere and covered by Ti and Au
overlayers Ni and Au sectors of the rotating disc, have been calculated analytically.\cite{27}
Then the constraints on $g_{an}$ for different $m_a$ were found from the inequality
\begin{equation}
\left|F_a^{\rm Ni,Ni}(d)-F_a^{\rm Ni,Au}(d)\right|\leq\Delta F_{\rm diff},
\label{eq3}
\end{equation}
\noindent
where predictions of the Lifshitz theory for the difference of Casimir forces
with the relaxation properties of free electrons disregarded were confirmed to
within $\Delta F_{\rm diff}=1~$fN error.
\begin{figure}[b]
\vspace*{-12cm}
\centerline{\hspace*{2cm}\includegraphics[width=6.50in]{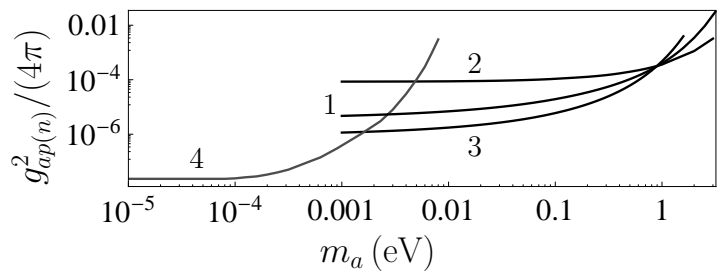}}
\vspace*{-7.9cm}
\caption{Constraints on the axion-to-nucleon coupling constant obtained from
measuring the difference of Casimir forces (line~1), the Casimir pressure (line~2)
and from the Casimir-less (line~3) and Cavendish-type (line~4) experiments are shown
as functions of the axion mass (see text for further discussion).
\protect\label{fg1}}
\end{figure}

The obtained constraints are shown by the line~1 in Fig.~\ref{fg1}.
For comparison purposes line~2 in the same figure indicates the constraints following
from measuring the Casimir pressure\cite{15} and line~3 the constraints found\cite{17}
from the Casimir-less experiment.\cite{18} The constraints of line~4 are derived\cite{28}
from the Cavendish-type experiment.\cite{29}
For all lines, the regions of the plane above the line are excluded by the results of
corresponding experiment and below a line are allowed.
As it follows from Fig.~\ref{fg1}, the constraints of line~1 are up to a factor of 14.6
stronger than the ones following from measuring the Casimir pressure (line~2) and, thus,
stronger than all other constraints derived from measurements of the Casimir
interaction.\cite{19} Only the line~3 found\cite{17}
from the Casimir-less experiment\cite{18} gives up to a factor of 2 stronger constraints.
This allows to conclude that the new constraints of line~1 are in good agreement with other
constraints obtained from previously performed experiments.

In fact the potentialities of experiments on measuring the Casimir interaction for obtaining
stronger constraints on axion-to-nucleon coupling constants are not exhausted. All these
experiments were not intended for constraining hypothetical interactions.
It is shown\cite{30} that minor modifications in the experimental setups on measuring
the lateral and normal Casimir forces between sinusoidally corrugated surfaces and in the
Casimir-less experiment would allow to further strengthen the obtained constraints over
wider regions of axion masses than in original experiments. It was also suggested to use
polarized test bodies.\cite{30a}

Quite recently, an improved experiment on measuring the Casimir pressure between two
parallel metallic plates at separations from 3 to $15~\mu$m has been proposed.\cite{31}
The suggested modifications in already existing experimental setup (the Cannex test of
quantum vacuum\cite{32}) allow more exact measurement of several important parameters,
including the separation distance between the plates, applied voltages, frequency shift
of the plate oscillation due to the Casimir pressure, vibration amplitude, characteristics
of patch potentials, and relative tilt angle of the plates. It is shown\cite{31} that
with the proposed improvements the Cannex test will be capable to directly measure
the thermal effect in the Casimir interaction and become a useful tool for investigation
of the dark matter and dark energy.

\section{Dark energy, cosmological constant and the quantum vacuum}

As mentioned in Sec.~1, the dark energy responsible for an accelerated expansion of the
Universe is well described by the cosmological constant in Einstein's equations.
In so doing, the value of dark energy density
$\varepsilon_{\rm de}\sim 10^{-9}~\mbox{J/m}^3$ required to explain the observed
acceleration corresponds to the cosmological constant
\begin{equation}
\Lambda=8\pi G \varepsilon_{\rm de}\approx 2\times 10^{-52}~\mbox{m}^{-2},
\label{eq4}
\end{equation}
\noindent
where $G$ is the gravitational constant.

It has long been noticed\cite{33} that the cosmological constant can be considered as
originating from the vacuum energy density of quantized fields.
This, however, leads to a problem that was named the vacuum catastrophe.\cite{5}
The point is that if one assumes the validity of local quantum field theory up to
the Planckian energy $E_{\rm Pl}\sim 10^{19}~\mbox{GeV}\sim 10^9~$J
and makes a cutoff in the divergent vacuum energy at the Planckian momentum,
the obtained energy density of order $10^{111}~\mbox{J/m}^3$ exceeds the above
value of $\varepsilon_{\rm de}$ by the factor $10^{120}$ (see Refs.~\citen{1,34}
for a review).

It was argued,\cite{35} however, that the renormalization procedure of quantum field
theory consisting in a transition from the nonobservable (bare) to physical values
of different quantities may provide a plausible explanation for the enormous difference
between the values of  $\varepsilon_{\rm de}$ and the energy density of quantum vacuum.
Actually, the divergent vacuum-vacuum expectation
values of the stress-energy tensor of $P$ bosonic fields with masses
$m_1,\,m_2,\,\ldots,\,m_P$ and $g_1,\,g_2,\,\ldots,\,g_P$ degrees of freedom and
$Q$ fermionic fields with masses
$M_1,\,M_2,\,\ldots,\,M_Q$ and $h_1,\,h_2,\,\ldots,\,h_Q$ degrees of freedom,
which are usually dropped by means of the normal ordering procedure of creation and
annihilation operators, are given by
\begin{equation}
\langle0|T_{ij}(x)|0\rangle=\frac{1}{2(2\pi)^3}\int d^3pp_ip_j
\left(\sum_{l=1}^{P}\frac{g_l}{\sqrt{m_l^2+\mbox{\boldmath$p$}^2}}-
\sum_{l=1}^{Q}\frac{h_l}{\sqrt{M_l^2+\mbox{\boldmath$p$}^2}}\right).
\label{eq5}
\end{equation}

Applying the method of dimensional regularization, it is easy to rewrite Eq.~(\ref{eq5})
in the space-time of $(4+2\epsilon)$-dimensions where $\epsilon$ is a complex number.
Then, by considering $\epsilon\to 0$, one obtains\cite{35}
\begin{equation}
\langle0|T_{ij}(x)|0\rangle=I_{\epsilon}g_{ij},
\label{eq6}
\end{equation}
\noindent
where
\begin{eqnarray}
&&I_{\epsilon}=\frac{1}{64\pi^2}\left[\sum_{l=1}^{P}g_lm_l^4\left(
\frac{1}{\epsilon}-\frac{3-2\gamma}{2}+
\ln\frac{m_l^2}{4\pi m_f^2}\right)\right.
\nonumber \\
&&~~~~~~~~~~~~~~~
\left.-\sum_{l=1}^{Q}h_lM_l^4\left(
\frac{1}{\epsilon}-\frac{3-2\gamma}{2}+
\ln\frac{M_l^2}{4\pi m_f^2}\right)\right],
\label{eq7}
\end{eqnarray}
\noindent
$\gamma\approx0.577$ is the Euler constant and  $m_f$ is a fictitious mass introduced
to preserve the same dimension of the stress-energy tensor as in 4-dimensional
space-time.

{}From Eq.~(\ref{eq6}) it becomes clear that subtraction of the quantity (\ref{eq5}) in order
to make equal to zero the stress-energy tensor of the quantum vacuum in Minkowski space-time
is equivalent to the renormalization of the bare cosmological constant
$\Lambda_{\epsilon}^{(b)}$ with its physical (renormalized) value $\Lambda^{\!\rm(ren)}$
equal to zero.
In curved space-time the value of
\begin{equation}
\Lambda^{\!\rm(ren)}=\Lambda=\Lambda_{\epsilon}^{(b)}+
8\pi G I_{\epsilon}\approx 2\times 10^{-52}~\mbox{m}^{-2}
\label{eq8}
\end{equation}
\noindent
is determined from the observed acceleration of the Universe expansion.
As a result, only the finite value $\Lambda^{\!\rm(ren)}$ enters the Einstein equations
after renormalization.

Within this approach, an infinitely large bare cosmological constant determined by the
00-component of the stress-energy tensor (\ref{eq5}) and (\ref{eq6}) should be viewed
as of little physical importance. Specifically, the infinitely large energy
density of the quantum vacuum should not be considered as a source of the gravitational
field as well as the bare charge in quantum electrodynamics is not a source of physical
Coulomb interaction. The source of physical gravitational interaction is described by
$\Lambda^{\!\rm(ren)}$ and results in the finite energy density $\varepsilon_{\rm de}$.
This is in close analogy to the Casimir effect where the finite (Casimir) energy
density and pressure in a quantization region restricted by some material boundaries
are obtained after subtraction of the infinitely large stress-energy tensor (\ref{eq5})
defined in an unbounded Minkowski space-time. In doing so, only the resulting finite
and measured in many experiments energy density is the source of gravitational
interaction.\cite{36}

\section{Conclusions and discussion}

In the foregoing, we have considered some new results on constraining the
parameters of axion-like particles as the most probable constituents of
dark matter. It was noted that recent experiment on measuring the
difference of Casimir forces not only leads to competitive constraints on
axion-to-nucleon interaction, but provides stronger support to the
constraints obtained from previous measurements of the Casimir interaction.
One can conclude that the Casimir physics has potentialities for deriving
even stronger constraints on the  axion-to-nucleon interaction.

Concerning the problem of dark energy and the concept of cosmological
constant, it was argued that a contradiction between the observed values
and theoretical predictions of quantum field theory is probably exaggerated.
Enormously or even infinitely large energy density and pressure of the
quantum vacuum and related values of bare cosmological constant should not
be treated as catastrophic because only the experimentally determined
renormalized quantities are of physical significance like it is the case
in QED and other quantum field theories
 of the Standard Model.
The existence of vacuum condensates does not necessarily mean large energy
density of the vacuum due to smallness of respective masses.
In this regard, an enormously large
energy density of the quantum vacuum cannot not be considered as a
source of gravitational interaction keeping in mind that only the
density of dark energy related to a renormalized cosmological constant
should gravitate. The above argumentation treating the cosmological
constant as one more fundamental constant of nature may appear somewhat
incomplete in the absence of conclusive quantum theory of gravitation.
It seems, however, that already developed quantum field theory in curved
space-time\cite{37,38} provides enough reason in favor of this approach.

\section*{Acknowledgments}

The author was partially funded by the Russian Foundation for Basic
Research, Grant No. 19-02-00453 A. His work was also partially supported by the
Russian Government Program of Competitive Growth of Kazan Federal University.
Helpful discussions with G. L. Klimchitskaya are acknowledged.


\begin{thebibliography}{99}
\bibitem{1}
J. A. Frieman, M. S. Turner and  D. Huterer, {\it Annu. Rev. Astron.
Astrophys.} {\bf 46}, 385 (2008).
\bibitem{2}
V. Trimble, {\it Annu. Rev. Astron. Astrophys.} {\bf 25}, 425 (1987).
\bibitem{3}
I. G. Irastorza and J. Redondo, {\it Progr. Part. Nucl. Phys.} {\bf 102},
89 (2018).
\bibitem{4}
P. J. E. Peebles and B. Ratra, {\it Rev. Mod. Phys.} {\bf 75}, 559 (2003).
\bibitem{5}
R. J. Adler, B. Casey and O. C. Jacob, {\it Amer. J. Phys.}
{\bf 63}, 620 (1995).
\bibitem{6}
M. S. Safronova, D. Budker, D. DeMille, D. F. J. Kimball, A. Derevianko
and C. W. Clark, {\it Rev. Mod. Phys.} {\bf 90}, 025008 (2018).
\bibitem{7}
G. Raffelt, {\it Phys. Rev. D} {\bf 86}, 015001 (2012).
\bibitem{8}
W. C. Haxton and K. Y. Lee, {\it Phys. Rev. Lett.} {\bf 66}, 2557 (1991).
\bibitem{9}
J. E. Kim, {\it Phys. Rep.} {\bf 150}, 1 (1987).
\bibitem{10}
E. G. Adelberger, E. Fischbach, D. E. Krause and R. D. Newman,
{\it Phys. Rev. D} {\bf 68}, 062002 (2003).
\bibitem{11}
S. Aldaihan, D. E. Krause, J. C. Long and W. M. Snow,
{\it Phys. Rev. D} {\bf 95}, 096005 (2017).
\bibitem{12}
M. Bordag, G. L. Klimchitskaya, U. Mohideen and V. M. Mostepanenko,
{\it Advances in the Casimir Effect} (Oxford University Press,
Oxford, 2015).
\bibitem{13}
V. B. Bezerra, G. L. Klimchitskaya, V. M. Mostepanenko and C. Romero,
{\it Phys. Rev. D} {\bf 89}, 035010 (2014).
\bibitem{14}
V. B. Bezerra, G. L. Klimchitskaya, V. M. Mostepanenko and C. Romero,
{\it Phys. Rev. D} {\bf 89}, 075002 (2014).
\bibitem{15}
V. B. Bezerra, G. L. Klimchitskaya, V. M. Mostepanenko and C. Romero,
{\it Eur. Phys. J. C} {\bf 74}, 2859 (2014).
\bibitem{16}
V. B. Bezerra, G. L. Klimchitskaya, V. M. Mostepanenko and C. Romero,
{\it Phys. Rev. D} {\bf 90}, 055013 (2014).
\bibitem{17}
G. L. Klimchitskaya and V. M. Mostepanenko, {\it Eur. Phys. J. C}
{\bf 75}, 164 (2015).
\bibitem{18}
Y.-J. Chen, W. K. Tham, D. E. Krause, D. L\'{o}pez, E. Fischbach
and R. S. Decca, {\it Phys. Rev. Lett.} {\bf 116}, 221102 (2016).
\bibitem{19}
V. M. Mostepanenko, {\it Int. J. Mod. Phys. A} {\bf 31}, 1641020 (2016).
\bibitem{20}
G. L. Klimchitskaya and V. M. Mostepanenko, {\it Contemp. Phys.}
{\bf 47}, 131 (2006).
\bibitem{21}
C.-C. Chang, A. A. Banishev, R. Castillo-Garza, G. L. Klimchitskaya,
V. M. Mostepanenko and U. Mohideen,
{\it Phys. Rev. B} {\bf 85}, 165443 (2012).
\bibitem{22}
A. A. Banishev, C.-C. Chang, G. L. Klimchitskaya, V. M. Mostepanenko
and U. Mohideen, {\it Phys. Rev. B} {\bf 85}, 195422 (2012).
\bibitem{23}
A. A. Banishev, G. L. Klimchitskaya, V. M. Mostepanenko and U. Mohideen,
{\it Phys. Rev. Lett.} {\bf 110}, 137401 (2013).
\bibitem{24}
A. A. Banishev, G. L. Klimchitskaya, V. M. Mostepanenko and U. Mohideen,
{\it Phys. Rev. B} {\bf 88}, 155410 (2013).
\bibitem{25}
G. Bimonte, D. L\'{o}pez and R. S. Decca, {\it Phys. Rev. B} {\bf 93},
184434 (2016).
\bibitem{26}
J. Xu, G. L. Klimchitskaya, V. M. Mostepanenko and U. Mohideen,
{\it Phys. Rev. A} {\bf 97}, 032501 (2018).
\bibitem{27}
G. L. Klimchitskaya and V. M. Mostepanenko, {\it Phys. Rev. D}
{\bf 95}, 123013 (2017).
\bibitem{28}
E. G. Adelberger, B. R. Heckel, S. Hoedl, C. D. Hoyle, D. J. Kapner
and A. Upadhye, {\it Phys. Rev. Lett.} {\bf 98}, 131104 (2007).
\bibitem{29}
D. J. Kapner, T. S. Cook, E. G. Adelberger, J. H. Gundlach, B. R. Heckel,
C. D. Hoyle and H. E. Swanson, {\it Phys. Rev. Lett.} {\bf 98}, 021101
(2007).
\bibitem{30}
G. L. Klimchitskaya, {\it Eur. Phys. J. C} {\bf 77}, 315 (2017).
\bibitem{30a}
V. B. Bezerra, G. L. Klimchitskaya, V. M. Mostepanenko and C. Romero,
{\it Phys. Rev. D} {\bf 94}, 035011 (2016).
\bibitem{31}
G. L. Klimchitskaya, V. M. Mostepanenko, R. I. P. Sedmik and
H. Abele, {\it Symmetry} {\bf 11}, 407 (2019).
\bibitem{32}
R. Sedmik and P. Brax, {\it J. Phys.: Conf. Series} {\bf 1138},
012014 (2018).
\bibitem{33}
Yu. B. Zel'dovich, {\it Sov. Phys. Usp.} {\bf 11}, 381 (1968).
\bibitem{34}
S. Weinberg, {\it Rev. Mod. Phys.} {\bf 61}, 1 (1989).
\bibitem{35}
V. M. Mostepanenko and G. L. Klimchitskaya, {\it Symmetry} {\bf 11},
314 (2019).
\bibitem{36}
G. Bimonte, E Calloni, G. Esposito and L. Rosa, {\it Phys. Rev. D}
{\bf 74}, 085011 (2006).
\bibitem{37}
A. A. Grib, S. G. Mamayev and V. M. Mostepanenko, {\it Vacuum
Quantum Effects in Strong Fields} (Friedmann Laboratory Publishing,
St.Petersburg, 1994).
\bibitem{38}
N. D. Birrell and C. P. W. Davies, {\it Quantum Fields in Curved
Space} (Cambridge University Press, Cambridge, 1982).
\end{thebibliography}
\end{document}